\documentclass[aps,prb,floatfix,superscriptaddress,twocolumn]{revtex4}
\usepackage{graphicx}
\usepackage{amsmath}
\usepackage{amssymb}
\usepackage{float}
\usepackage{bm}
\newcommand{\ek}{\epsilon_{\mathbf{k}}}

\begin{document}

\title{Pair-breaking effects in the Pseudogap Regime: \\Application to High Temperature Superconductors} 
\author{Ying-Jer Kao}
\email{y2kao@uwaterloo.ca}
\affiliation{Department of Physics, University of Waterloo, Waterloo, ON,
  N2L3G1, Canada} 
\author{ Andrew P. Iyengar}
\author{ Jelena Stajic}
\author{K. Levin}
\affiliation{James Franck Institute and Department of Physics, University
  of Chicago, Chicago, IL 60637, U.S.A.}
\date{\today}
\begin{abstract}
Abrikosov-Gor'kov (AG) theory, the foundation for understanding
pair-breaking effects in conventional superconductors,
is inadequate when there is an excitation gap (pseudogap)
present at the onset of superconductivity. In this paper we
present an extension of AG theory within two important,
and diametrically opposite approaches
to the cuprate pseudogap. The effects of impurities on
the pseudogap onset temperature $T^*$ and on $T_c$, 
along with comparisons to  
experiment are addressed.
\end{abstract}


% Transition temperature variations
% Theories and models of superconducting state,
% General properties; correlations between physical properties in normal and superconducting state, 
% Effects of crystal defects, doping and substitution 

\maketitle

% Motivation

Impurity effects in the high temperature superconductors
have been the subject of a large body of 
experimental and theoretical literature concentrating
on pair-breaking effects on $T_c$\cite{Radtke1992,Franz1997,Tallon1997,Williams1998}, 
$d$-wave density of states effects
near $T \approx 0$,
\cite{Lee1,Fisher,Hirschfeld}, 
local suppressions of the order parameter,
\cite{Walker,Davis},
transport effects,
\cite{Lee2}, 
and aspects of the
superconductor-insulator transition
\cite{Uchida}. 
Although there are works on the effects of a single impurity in the
pseudogap models\cite{Balatsky,Carbotte2,Wang,Morr}, 
with very few exceptions\cite{Williams1998,Qijin} 
little theoretical
attention has been paid to the interplay between the widely observed
cuprate pseudogap and the effects of disorder on pair-breaking. 
This is a particularly striking omission, given that a major
fraction of the superconducting phase diagram\cite{TallonReview} is associated
with a pseudogap. 
The goal of the present paper is to establish a formal 
mean-field structure
(analogous to Abrikosov-Gor'kov (AG) theory) that incorporates 
this pseudogap in computing both $T_c$ and gap onset temperature, $T^*$,
along with other derived properties.  Here we
address two 
mean-field approaches ( orthogonal in their physics, but
similar in their formalism), to the incorporation
of the pseudogap: one in which the pseudogap  derives from superconductivity
itself\cite{Kosztin1,Chen2,Janko} 
(``intrinsic") and one in which it is ``extrinsic",
either associated with a hidden order parameter\cite{Laughlin,DiCastro},
or with band-structure effects\cite{Loram,Nozieres2}. 
This intrinsic pseudogap\cite{Kosztin1,Chen2}
arises from
a stronger than BCS attractive interaction which leads to
finite momentum pair excitations of the normal state
and condensate.

In contrast to BCS theory, in the pseudogap
phase there is an excitation gap \textit{present at $T_c$},
which, at low doping $x$, remains
relatively $T$-independent for all $T \le T_c$\cite{Norman}. This
necessarily will affect 
(i) fundamental characteristics of the 
superconducting phase as well as (ii) the nature of impurity
pair-breaking. Indeed, to support (i),
there are strong indications
from thermodynamics\cite{TallonReview}
and tunneling\cite{Deutscher} experiments
that the effects of the normal state pseudogap persist below $T_c$\cite{Chen2001}. 
%This persistence will be addressed here in the context of the superfluid
%density.
% which we address here.
Evidence in support of (ii) comes
from the fact that pseudogap
effects appear to correlate with the degree of the $T_c$
suppression in the presence of Zn impurities\cite{Tallon1997}.
This suppression becomes progressively more rapid as the
size of the pseudogap grows.

Both intrinsic and extrinsic models for the pseudogap
are associated with a generic set of mean-field equations.
% in which differences arise only via the dispersion of fermions.
It is reasonable to stop at
a mean-field level because these materials (in some,
but not all respects) do not appear
to be strikingly different from BCS superconductors, and because
the true critical regime appears to be rather narrow\cite{Larkin}. Moreover,
we believe fluctuation effects around strict BCS theory such
as the phase fluctuation model of Emery and co-workers\cite{Emery} are unlikely 
to explain the often very large separation observed between the gap
onset temperature $T^*$ and $T_c$. It seems more appropriate, thus
to search for an improved mean field theory\cite{Larkin}.
Then additional fluctuation effects can be appended as needed.

In this generalized mean field approach, in the clean limit 
and for $T \le T_c$, the gap and number equations are given by
\begin{subequations}
\label{eq:self-consistent}
\begin{equation}
1+ g_{sc}\, T\sum_{n, \mathbf{k},\alpha} \frac{\varphi^2_{{\bf k}}}{{\omega}_n^2+ 
{E}^{\alpha\; 2}_{\mathbf{k}}
}=0,\label{eq:gap}
\end{equation}
\begin{equation}
   n = \frac{1}{ 2} - T \sum_{n,{\mathbf k},\alpha}  
                 \frac{i {\omega}_n
                   +\epsilon_{\bf k}^\alpha - {\mu}}
                 {{\omega}_n^2+
                   {E}^{\alpha\; 2}_{\mathbf{k}} },\label{eq:number}
\end{equation}
\end{subequations}
where $g_{sc}$ is the coupling constant for the superconducting order
parameter, $\varphi_{{\bf k}}=(\cos k_x-\cos k_y)$ is the $d$-wave symmetry factor,
${\Delta}_{sc}$ represents the superconducting order parameter,
and ${\Delta}_{pg}$ the pseudogap which persists
in the $T \le T_c $ phase. Finally, $\alpha$
is a band index, which appears
in  some microscopic approaches\cite{Laughlin,DiCastro}  to the extrinsic case. The momentum summation in the
extrinsic case is over the reduced Brillouin zone.
%The ``tilde" on various symbols here and throughout the paper,
%corresponds to the appropriate renormalization associated with
%impurities. 
These equations depend in an important
way on the electronic dispersion which differs
in the two schemes. In the intrinsic school the fermionic dispersion is  
characterized by
\begin{subequations}
\label{eq:in-disp}
\begin{eqnarray}
{E}_{\bf k}^2 &=& (\ek - {\mu})^2 + {\Delta}^2(T), \\
{\Delta}^2(T) &=& {\Delta}_{pg}^2 (T) +{\Delta}_{sc}^2 (T),\label{eq:fullgap} \\
\ek &=& \xi_{\bf k}.
\end{eqnarray}
\end{subequations}
Here $\xi_{\mathbf{k}}$ is the ``bare" band structure, taken to 
correspond to a nearest neighbor tight-binding model.
This should be contrasted
with that in the extrinsic school,
\begin{subequations}
\label{eq:ex-disp}
\begin{eqnarray}
{E}_{\bf k}^{\pm\; 2} 
&=& (\ek^\pm - {\mu})^2 + {\Delta}_{sc}^2 (T), \\
\ek^\pm &=& \pm \sqrt{\xi_{\bf k}^2 + {\Delta}_{pg}^2(T) }.
\end{eqnarray}
\end{subequations}
%%The $1/2$ in Eq.~\eqref{eq:Density} comes from the
%%proper handling of the convergence factor\cite{Schafroth1997}.
%Note a subtle but important
%distinction between the two cases through the placement
%of the chemical potential. 
%%For this extrinsic case, only in the limit ${\mu} = 0$, can
%%one readily define a conventional excitation gap ${\Delta}$ 
%%which, incidentally then satisfies Eq.~\eqref{eq:fullgap}.
%This reflects a
%fundamental difference between the two theories: in the intrinsic
%case the pseudogap enters the fermion velocity and dispersion through 
%particle-hole mixing of the fermionic quasi-particles, 
%whereas in the extrinsic case it reflects particle-particle mixing. 
The fermionic dispersions of the two schools differ as a direct 
consequence of the mechanisms that generate the respective pseudogaps. 
At the mean field level, a pseudogap due to pairing
correlations forms as particles and holes mix to form the 
fermionic quasiparticles.
Those of a spin- or charge- ordered state, though, are
particle-particle mixtures. 
In the regime $T \le T_c$, where sharp excitations exist, 
these can be taken as the defining
characteristics of ``intrinsic'' and ``extrinsic'' models of the pseudogap.
Since generally $\mu \ne 0$ away from half-filling, 
only in the intrinsic school is the pseudogap 
pinned at the Fermi surface.

The respective properties of the pseudogap lead to equations for 
its magnitude, which we summarize for
$T \le T_c$, in terms of the particle-particle ($\chi^{pp}$)
and particle-hole ($\chi^{ph}$) susceptibilities.
For the intrinsic school 
\begin{eqnarray} 
\chi^{pp}(\mathbf{q},i\omega_n) &=& 
T\mathop{\sum_{\mathbf{k},m}} 
\frac{i\nu_m + \epsilon_{\mathbf{k}}}{\nu_m^2 + E_{\mathbf{k}}^2}
\frac{\varphi^2_{\mathbf{k} - \mathbf{q}/2}}
{i(\nu_m - \omega_n) + \epsilon_{\mathbf{k} - \mathbf{q}}}, \nonumber\\
{\Delta}_{pg}^2 &=& -T\sum_n\sum_{\mathbf{q}\ne 0} 
\frac{g_{sc}}{1+g_{sc} {\chi}^{pp}(\mathbf{q},i\omega_n, \Delta)}. 
\label{eq:pgeqn}
\end{eqnarray}

Note that ${\chi}^{pp}$ depends\cite{Kosztin1,Janko} on the full excitation
gap $\Delta$. Here
 $\Delta_{pg}(T)$ is associated with the number of finite
momentum pair excitations of the condensate. These occur when the
strength of the
attractive interaction $g_{sc}$ is progressively increased, so that it
is larger than that associated with the BCS regime.
For the extrinsic school, the counterpart equation is
\begin{eqnarray}
{\chi}^{ph}(\mathbf{0},0,\Delta_{sc},\Delta_{pg}) 
&=& 
T\sum_{n, \mathbf{k},\alpha} 
\frac{\varphi^2_{{\bf k}}(\epsilon_{\bf k}^{\alpha}-\mu)}
{({\omega}_n^2+{E}^{\alpha\; 2}_{\mathbf{k}}) \epsilon _{\bf k} ^{\alpha}}\nonumber\\
&=& -g_{pg}^{-1} ,
\label{eq:pgeqn2}
\end{eqnarray}
where $g_{pg}$ is the coupling constant for the pseudogap order and
the momentum summation is over half of the Brillouin zone. Here we consider
the pseudogap with same $d$-wave structure as the superconducting order. 

\begin{figure}
\includegraphics[width=3.4in,clip]{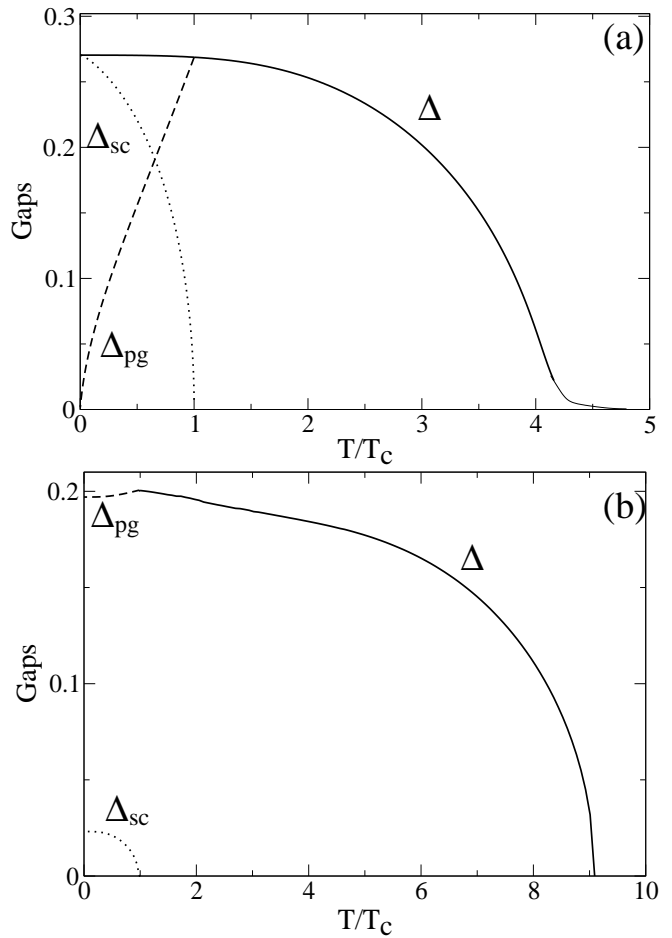}
\caption{Energy gaps for intrinsic (a) and extrinsic (b) cases. Solid lines
  are the total excitation gaps, dotted lines  the superconducting order
  parameters and dashed lines the pseudogaps below $T_c$. The gaps are in
  units of $4t_\parallel$. The curve
for $T \ge T_c$ in (a) represents a rough extrapolation. }
\label{fig:gaps}
\end{figure}

Figure~\ref{fig:gaps} shows the temperature dependencies of the 
different energy gaps obtained
by solving the complete set of equations in the two pseudogap schools 
within
the underdoped regime.
In the intrinsic case $T^*$ marks the gradual onset of the
pseudogap, which is associated with bosonic or
pair excitations formed in the presence of
a stronger-than-BCS attractive interaction. Only at and below $T_c$
does the identification of $\Delta$ become precise, so that for
this (intrinsic) case we plot an extrapolation of Eqs. \ref{eq:gap}, 
\ref{eq:number}, and \ref{eq:pgeqn} to $T \ge T_c$.
Figure~\ref{fig:gaps}a shows
that below $T_c$ the fraction of the bosonic population joining
the condensate of zero-momentum pairs ($ \propto \Delta_{sc}^2$)
increases at the expense of the finite-momentum bosonic fraction
($\propto \Delta_{pg}^2$) until the fully condensed ground state
is reached.   
By contrast, for the extrinsic case (Fig.~\ref{fig:gaps}b) 
superconductivity forms on
top of a pre-existing excitation gap in the
effective band structure which first appears at $T^*$, the phase transition
temperature marking the onset of the extrinsic order.

One can capture the key physics of these two schemes in a reasonably
accurate
phenomenological approach.
The bosonic excitations associated with the mean-field theory\cite{Kosztin1} of 
Eqs.~\eqref{eq:gap}, \eqref{eq:number} and \eqref{eq:pgeqn} lead to the
temperature dependence of the pseudogap below the clean limit critical
temperature $T_{c0}$ 
\begin{equation}
{\Delta}_{pg}^2 (T) \approx {\Delta}^2 (T_{c0}) 
\left(\frac{T}{T_{c0}}\right)^{3/2},\quad T \le T_{c0}.
\label{eq:in-pg}
\end{equation}
These bosons are, thus, associated with a quasi-ideal Bose gas.
By contrast for the extrinsic case, in the well-established
pseudogap regime, below $T_{c0}$, the pseudogap is relatively
$T$-independent
\begin{equation}
{\Delta}_{pg}^2 (T)  = {\Delta}^2 (T_{c0}), \quad T \le T_{c0}.
\label{eq:ex-pg}
\end{equation}
Here we define $\Delta (T_{c0}) = \Delta_{pg} (T_{c0})$.
In both Eqs.~\eqref{eq:in-pg} and \eqref{eq:ex-pg} above,
 we may view ${\Delta}(T_{c0})$ as a phenomenological parameter 
taken from experiment\cite{TallonReview}.
We will adopt this approach here in large part because it provides
a more readily accessible theoretical framework for the
community, 
and because it
connects more directly with experiment.

\begin{figure}
\includegraphics[width=3.2in,clip]{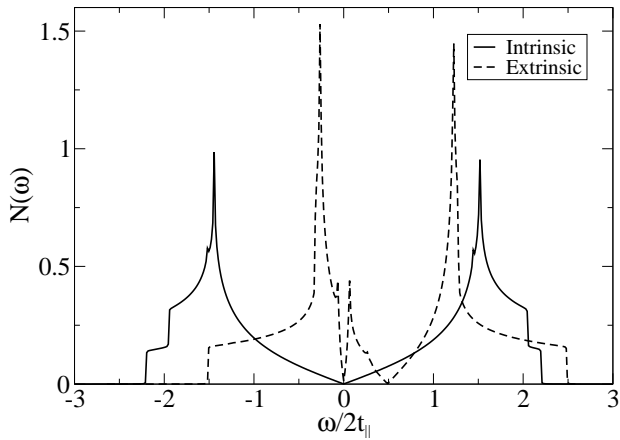}
\caption{ DOS for intrinsic and extrinsic models at $T=0$. Only one gap structure
  appears in the intrinsic DOS, while two distinct gap structures appear
  in the extrinsic DOS. }
\label{fig:cleanDOS}
\end{figure}

The pronounced differences between the fermionic
dispersion in these
two theoretical schools can be seen from the associated densities of
states (DOS) plotted in 
Fig.~\ref{fig:cleanDOS}, which compares the 
intrinsic and extrinsic models at $T=0$.  
In the intrinsic model one sees
only one excitation gap feature\cite{Lifetime}
$\Delta=\sqrt{\Delta_{sc}^2+\Delta_{pg}^2}$
in Fig.~\ref{fig:cleanDOS}, centered around the Fermi energy.
Van Hove singularities are also apparent here as relatively sharp
structures.
In contrast, 
there exist two distinct features for the extrinsic theory. 
The more prominent pseudogap peaks are centered around $-\mu$, while the
superconducting peaks appear around the Fermi energy\cite{Carbotte1}. 
Indeed, for this extrinsic case, only in the limit
$\mu = 0$ can one readily define an excitation gap
$\Delta$ as in a conventional 
superconducting phase\cite{Carbotte1}, satisfying
Eq.~\eqref{eq:fullgap}.
That the superconducting order parameter and pseudogap
contribute to separate features in the density of states
represents a rather clear signature of this extrinsic pseudogap
school.  To date, the bulk of experimental tunneling data
supports a picture in which there is a single excitation gap 
feature\cite{Renner1998b,Zasadzinski}, although there are some reports
of multiple gap structures
in c-axis intrinsic tunneling spectroscopy\cite{Krasnov2000}.
At $T=T_c$, the extrinsic superconducting gap closes and the densities
of states for the two schools become quite similar, save for the pinning
of the gap minimum to the Fermi surface in the intrinsic case.

%WHY ARE PG PEAKS HIGHER? WHY IS THE BAND WIDTH FOR INTRINSIC CASE LARGER
% THAN 8t( $t_\perp$ EFFECT)?   HOW IS $\mu$ MEASURED, WHY NEGATIVE IN THE INTRINSIC CASE? 
%%DISCUSS VAN HOVE

We turn now to impurity effects which, just as in the BCS
case, are not expected to change the formal 
structure of our mean field theory. The greatest
complication is associated with the impurity-renormalized 
$\tilde{\Delta}_{pg}$,
calculated from all possible diagrammatic insertions of the impurity vertex
into the particle-hole and particle-particle susceptibilities
[see Eqs.~\eqref{eq:pgeqn} and \eqref{eq:pgeqn2}.] 
A detailed study of these
effects in the 
intrinsic case appears in
Ref.~\onlinecite{Qijin}, although here
we will proceed more phenomenologically within both schools.
We base the present treatment on analogs of the clean limit
mean field gap equations Eqs.~\eqref{eq:gap} and \eqref{eq:number}
with substitutions
$\Delta_{sc}\rightarrow \tilde{\Delta}_{sc}$, 
$\Delta_{pg}\rightarrow \tilde{\Delta}_{pg}$,
$\omega_n\rightarrow\tilde{\omega}_n$, and $\mu\rightarrow\tilde{\mu}$.
At the phenomenological level the $T$-dependence
of the intrinsic pseudogap is given by
\begin{equation}
\tilde{\Delta}_{pg}^2 (T) \approx \tilde{\Delta}^2 (T_c) 
\left(\frac{T}{T_c}\right)^{3/2},\quad T \le T_c,
\label{eq:in-pg-dirty}
\end{equation}
and for the extrinsic case, 
\begin{equation}
\tilde{\Delta}_{pg}^2 (T)  \approx \tilde{\Delta}^2 (T_c), \quad T \le T_c.
\label{eq:ex-pg-dirty}
\end{equation}
where, in both schools,
the excitation gap $\tilde{\Delta}(T_c)$, is presumed
to be determined from experiment.
 
To compute the renormalized frequency $i\tilde{\omega}(i\omega_n)$ 
and chemical potential $\tilde{\mu}(i\omega_n)$,
we follow the usual impurity $T$-matrix approach. 
We presume
an $s$-wave short-range impurity potential
$V({\bf r})=u \delta({\bf r}-{\bf r}_i)$. 
The impurity scattering matrix $\hat{T}(\omega_n)$ in Nambu space satisfies 
the Lippman-Schwinger equation:
$\hat{T}(\omega_n)=u\hat{\sigma}_3
\left( 1 + \hat{T}(\omega_n)  \sum_{\bf k} \hat{g}({\bf k},\omega_n)
\right)$,
where $\hat{g}$ is the impurity-dressed Green's function,
\begin{equation}
\hat{g}({\bf k}, i\omega_n) = 
\frac{i\tilde{\omega}_n\hat{\sigma}_0 
+ {\bm \Delta }({\bf k})\hat{\sigma}_1 
+ (\ek - \tilde{\mu})\hat{\sigma}_3}
{(i\tilde{\omega}_n)^2 - \tilde{E}_{\bf k}^2},
\end{equation}
Here ${\bm \Delta}$ is either the full gap or superconducting order parameter in the intrinsic
and extrinsic cases, respectively, and  $\hat{\sigma}_i$ are Pauli
matrices. Here we suppress the band index in the extrinsic case.
Labeling components as $\hat{g} = \sum_i g_i\hat{\sigma}_i$, 
the regular and anomalous
Green's functions are $\tilde{G} = g_0 + g_3$, 
$\tilde{F} = - g_1$.
The frequency and chemical potential are renormalized through impurity
self-energy $\hat{\Sigma}=n_i\hat{T}$, and 
$i\tilde{\omega}_n=i\omega_n-\Sigma_0,\quad \tilde{\mu}=\mu-\Sigma_3$, where
$n_i$ is the number of impurities per unit cell.   
We note that  the $T$-matrix for the extrinsic 
school depends only on the band structure and is independent 
of the specific type of extrinsic order.  

The components of the self-energy are given by 
\begin{equation}
  \Sigma_0=\frac{n_i g_0}{ (1/u-g_3)^2-g_0^2},\quad \Sigma_3=\frac{n_i
  (1/u-g_3)}{(1/u-g_3)^2-g_0^2},
\label{eq:dirtyselfenergy}
\end{equation}
%where $n_i$ is the number of impurities per unit cell 
and 
\begin{equation}
  g_0=\sum_{\bf k}\frac{i\tilde{\omega}_n}{(i\tilde{\omega}_n)^2- 
  \tilde{E}_{\bf k}^2},
\quad g_3=\sum_{\bf k}
\frac{\epsilon_{\bf k}-\tilde{\mu}}{(i\tilde{\omega}_n)^2-\tilde{E}_{\bf k}^2}.
\label{eq:dirtyselfconsistent}
\end{equation}
There is no frequency-dependent self-energy
associated with gap renormalization due to $d$-wave
symmetry.  
Finally, the magnitudes of $\tilde{\Delta}$,
$\tilde{\Delta}_{sc}$ and $\tilde{\Delta}_{pg}$ can be obtained
using Eqs.~\eqref{eq:self-consistent},\eqref{eq:in-disp} and \eqref{eq:in-pg}, presuming that
the excitation gap at $T_c$ is taken from experiment.
Here we take
the bare lattice dispersion $\xi_{\bf k}=-2 t_\parallel( \cos k_x + \cos
k_y)-2 t_\perp \cos k_z$ so that the
dimensionless  coupling constant is given by
$g/4t_\parallel$.

\begin{figure}
\includegraphics[width=3.2in,clip]{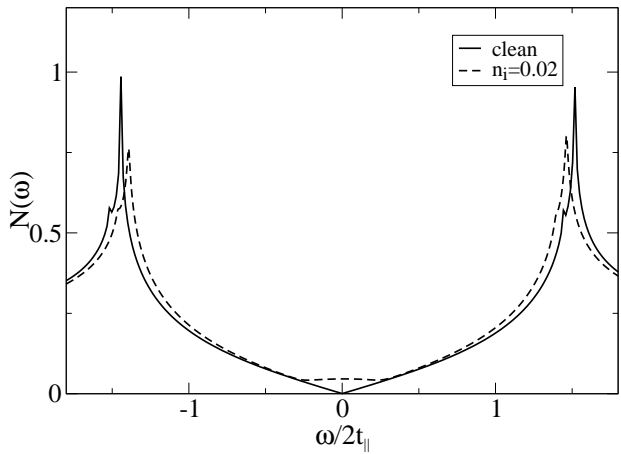}
\caption{Intrinsic DOS for the clean and dirty cases at $T=0$. The DOS is
  centered around the Fermi energy.}
\label{fig:intDOS}
\end{figure}

\begin{figure}
\includegraphics[width=3.2in,clip]{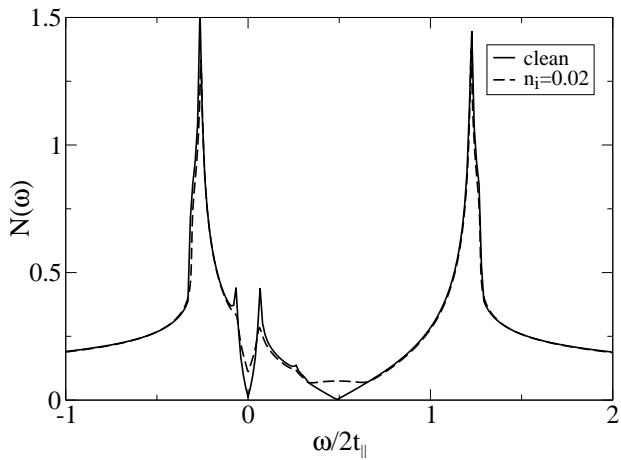}
\caption{Extrinsic DOS for the clean and dirty cases at $T=0$. The DOS is
  centered around $-\mu$.}
\label{fig:extDOS}
\end{figure}

\begin{figure}
\includegraphics[width=3.2in,clip]{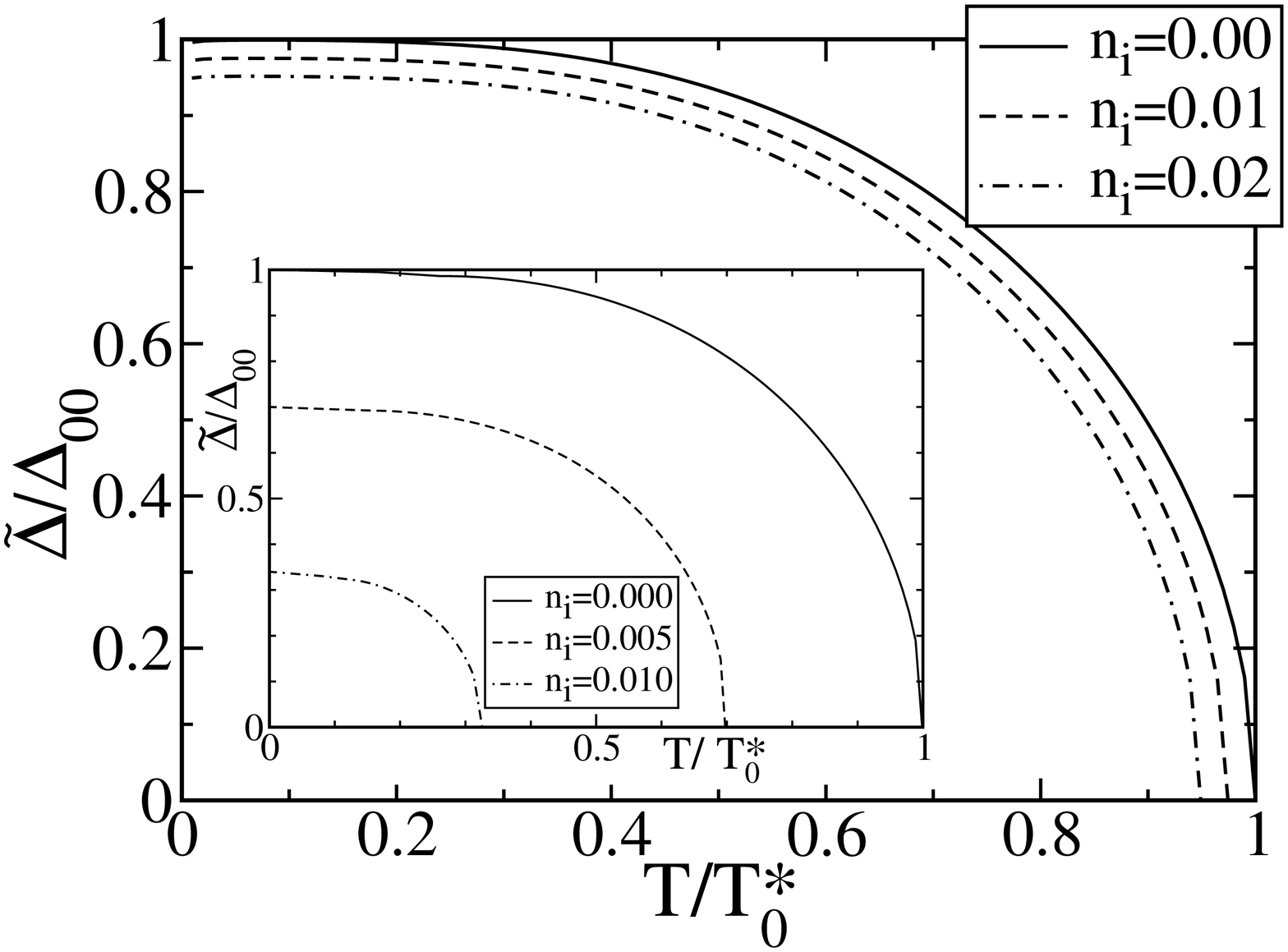}
\caption{Temperature dependencies of the full excitation gaps for the
  intrinsic case at  strong ($g/4t_\parallel=-1.2$) and weak
  coupling ($g/4t_\parallel=-0.15$, inset), in the
  unitary limit at
  different impurity densities. Temperatures are normalized to the clean
  limit $T^*_0$ and
  gaps are normalized to the zero-temperature values in the clean limit
  $\Delta_{00}$. 
%$T^*$ suppression due to the impurities can be read off
%  from the onset temperature of the gap.  
}
\label{fig:BtGap1}
\end{figure}

Figures ~\ref{fig:intDOS} and \ref{fig:extDOS} show the effects of
impurities (for unitary scattering) on the density of states
at $T=0$, in the intrinsic and extrinsic
cases respectively. As can be seen, particularly for the intrinsic
case, impurities decrease slightly the height and separation of the gap peaks [See Fig.~\ref{fig:BtGap1} below]
and fill in the low frequency region, but otherwise their
effects are not dramatic.  For the extrinsic school, the superconducting
gap region 
is more qualitatively affected by pair-breaking, while the pseudogap
peaks remain relatively robust. It can be inferred from these
figures that with increasing disorder the differences in the two
schools diminish, from the perspective of the
density of states, except that the position of the minimum
in the extrinsic case is not tied to the Fermi energy. 
Physical differences, however, remain profound,
particularly
in the electrodynamics\cite{Jelena,Drew} 
of the superconducting phase.

In the remainder of this paper we focus on the behavior of
$T_c$ (and $T^*$) and the appropriate generalization of
AG theory in the presence of a pseudogap.
%We address here only the intrinsic case,
%since once $\tilde{\Delta} (T_c))$ is specified the behavior of 
%the transition temperature is expected to be rather similar in the two schools.
For definiteness, we consider Zn doping experiments where
we exploit the experimental
observation that \textit{the excitation gap $\tilde{\Delta}$
at $T_c$ is relatively unaffected by Zn impurities}\cite{TallonReview}.
We focus here on  the unitary limit ($1/u=0$), which is regarded as relevant to
Zn doping in the cuprates. 
%Our calculations also address $T^*$, at a self consistent level
%for the intrinsic theory.  For the extrinsic school,
%one can infer from the constancy of $\tilde{\Delta} (T_c)$ that
%$T^*$ is relatively unaffected by these $Zn$ impurities.

We begin with the intrinsic school,
where the sensitivity of $T_c$ and $T^*$ to the impurity concentration $n_i$ 
can be studied as a function of a single coupling constant $g = g_{sc}$, 
which we presume to be unaffected by the addition of impurities.  
Figure~\ref{fig:BtGap1} shows the behavior of the excitation
gap $\tilde{\Delta}(T)$ vs temperature normalized to
its clean limit value, obtained using the impurity-generalized
form of Eqs.~\eqref{eq:gap} and
\eqref{eq:number}.
The figure should be viewed as extending above
$T_c$
only in the sense that it provides a reasonable extrapolation\cite{Maly2}
as well as estimate of $T^*$. 
In reality, Fig.~\ref{fig:gaps}a
indicates that
a \textit{crossover}
description for the excitation gap at $T$ above $T_c$ is more correct.
The main panel corresponds to 
 the  strong ($g/4t_\parallel=-1.2$) and the inset the
weak ($g/4t_\parallel=-0.15$)  coupling regimes for various
values of the impurity density $n_i$ in the unitary limit.
In the weak
coupling regime, the suppression of the gap is largest, as is the
reduction in $T^*$.
In the strong coupling case, the suppression is
smaller and at low impurity densities, the curves are very close to those
obtained in the clean limit, \textit{indicating smaller 
pair-breaking effects
on the excitation gap and its onset temperature
$T^*$}.

\begin{figure}
\includegraphics[width=3.4in,clip]{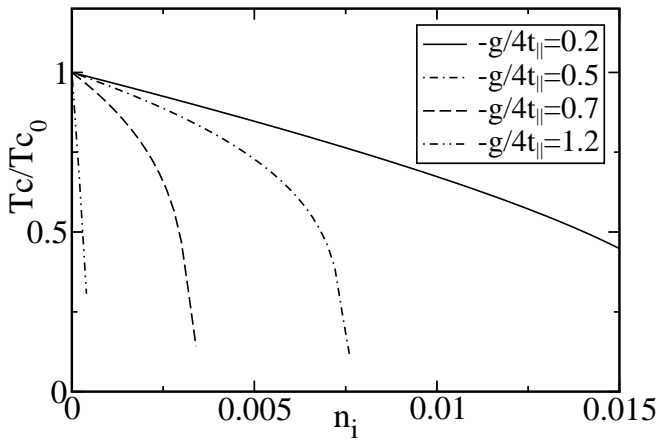}
\caption{$T_c$ suppression due to impurities for $1/u=0$ (unitary limit) in
  the intrinsic case. 
The temperatures are normalized to the clean limit $T_{c0}$.}
\label{fig:TcSuppress}
\end{figure}

Figure~\ref{fig:TcSuppress} shows the way in which impurities
suppress the phase coherence temperature $T_c$ 
at different coupling strengths (in the unitary limit),
based on the assumption, supported experimentally\cite{Tallon1997},
that the
excitation gap at the appropriate $T_c$ is relatively independent
of impurity concentration. It can be seen
that the suppression rate increases as the coupling
becomes stronger, or effectively
as $\tilde{\Delta}(T_c)$ increases. Similar results for the extrinsic case were
obtained in Ref.~\onlinecite{Williams1998}.
%However, it should be noticed that in the
%current calculations we self-consistently solved the chemical potential
%$\mu$ and took into account the renormalization of the energy dispersion
%due to the lack of particle-hole symmetry on a 2-D lattice.   
This faster $T_c$ suppression  in the strong coupling regime  can be 
understood through a simple physical picture. 
Impurity scattering will produce states which
fill in the gap and eventually destroy superconducting coherence. In the
strong coupling (pseudogap) regime, where the normal state already
has a gap,  fewer impurities are required to restore the system to
 the ``normal'' state.
%Thus the suppression rate is much larger when there exists
%a normal state gap at $T_c$.  

\begin{figure}
\includegraphics[width=3.2in,clip]{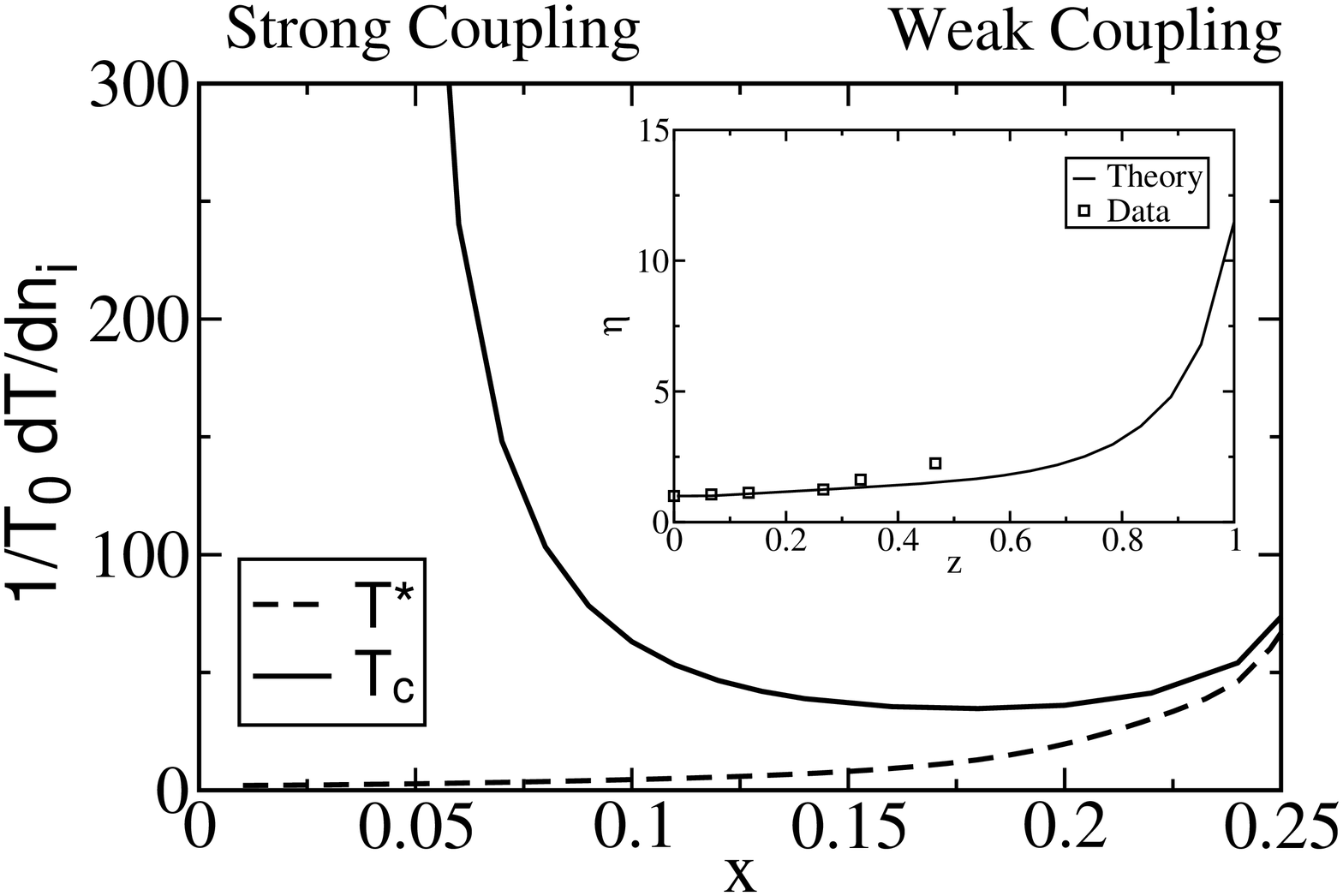}
\caption{Initial slopes of $T^*$ and $T_c$ suppression
  $\frac{1}{T_0}\frac{dT}{dn_i}$ \textit{vs} doping, in the  unitary  limit
for the intrinsic case.
The inset presents a comparison between theory and experimental data from Ref.~\onlinecite{Tallon1997}.
See text for details.
}
\label{fig:dT/dy-x}
\end{figure}

We turn now to calculations which can be directly compared with
experiment and plot the normalized slopes of $T^*$ and $T_c$ with respect
to increasing Zn concentration, for varying hole
concentration $x$, first for the intrinsic case. 
%We follow the procedure outlined in Ref.~\onlinecite{Chen2},
%assuming $g$ is independent of doping, while the bandwidth $4t_\parallel$
%decreases with underdoping, therefore the effective coupling
%$g/4t_\parallel$ becomes stronger with underdoping. 
To convert from the coupling constant parameter $g$ to $x$
we take as input the experimentally measured values of 
$\rho_s (x,0)$ and the measured excitation gap at $T_c$. Here it is
adequate to choose these values corresponding to the pristine case,
and presume that Zn doping does not affect the excitation gap
at $T_c$.
Figure~\ref{fig:dT/dy-x} indicates the initial slope ( $1/T_0 dT/dn_i$,
 where 
%$n_i$ is  the impurity concentration, and
 $T_0$ is the appropriate clean limit temperature) for 
$T^*$ (dashed line) or $T_c$ (solid line).
In the overdoped limit, 
the theory is asymptotically equivalent to
standard AG theory, in which also $T^* = T_c$.
%where it diverges as $1/T^*_0$.  
For smaller values of $x$ 
the slope decreases
so that  $T^*$ is only weakly dependent on impurity concentration.
By contrast, the initial $T_c$ slope (solid line) shows a very different hole
concentration dependence. 
As the hole concentration decreases, the slope decreases. However,
in the very underdoped regime, where the pseudogap is well established, 
the curve turns around and rapidly increases. 
The inset presents a comparison of theory and experiment\cite{Tallon1997}
as $\eta = (dT_c/dn_i)/(dT_c/dn_i, x=0.20)$ \textit{vs} $
z=\Delta_{pg}(T_c)/(\Delta_{pg}(T_c),x=0.05)$,
where the agreement appears to be reasonable.
There are fewer systematic studies of impurity-induced changes in
$T^*$; however, the small effect found here at low $x$ appears to be
compatible with the data.

\begin{figure}
\includegraphics[width=3.2in,clip]{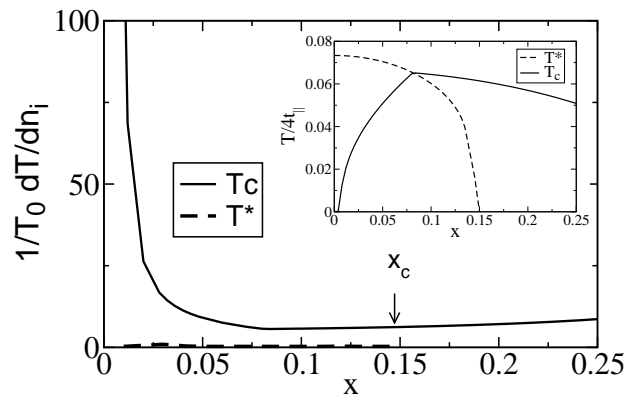}
\caption{Initial slopes of $T^*$ and $T_c$ suppression
  $\frac{1}{T_0}\frac{dT}{dn_i}$ \textit{vs} doping, in the  unitary  limit
for the extrinsic case. The inset shows the clean phase digram given by coupling
constants $g_{pg}/4t_\parallel = -0.4$ and
$g_{sc}/4t_\parallel =-0.375 $. The critical doping $x_c$ where
$\Delta_{pg}$ vanishes is around 0.15.  }
\label{fig:dT/dy-x-extrinsic}
\end{figure}

Finally in Fig.~\ref{fig:dT/dy-x-extrinsic} we present 
the counterpart plots of the initial
slopes for $T_c$ and $T^*$ in the extrinsic
case. $T_c$ is computed as in the intrinsic
case by assuming $\Delta_{pg}(T_c)$ is relatively insensitive to 
impurities. Impurity renormalizations are determined through
Eqs.~\eqref{eq:dirtyselfenergy} and \eqref{eq:dirtyselfconsistent}
while the suppression of $T^*$
is calculated via Eqs.~\eqref{eq:number} and \eqref{eq:pgeqn2},
extended to include appropriate impurity renormalizations. 
The inset shows the clean phase diagram which forms the basis for these
calculations. Our fit to the published form\cite{Carbotte2}
of this phase diagram provided values for the coupling constants $g_{pg}/4t_\parallel = -0.4$ and
$g_{sc}/4t_\parallel =-0.375 $.  To make contact with experiment we chose a
parameter set in which $T_c/T^*$ and $n_s/\Delta(0)$ were reasonably
well fit to experiment in the underdoped regime. As is similar
to the intrinsic case, there is a dramatic increase in the
slope of $T_c$ as the insulator is approached. This increase
is associated with the onset of the pseudogap which occurs for
$ x \le 0.15$. Above this critical concentration $T^*$ is zero,
and the system becomes a conventional dirty BCS superconductor.
In this way, the intrinsic and extrinsic schools differ, since
for the former at large $x$, $T^* \rightarrow T_c$. 
%In the highly overdoped case the impurity dependencies for $T^* = T_c$
%in the intrinsic case, and $T_c$ (only) in the extrinsic case
%are, then, governed
%by conventional AG
%theory.

The theoretical machinery that we have set up has strong
similarities to an approach taken by Loram and collaborators\cite{Loram},
extended further to the disordered case\cite{Tallon1997,Williams1998}.
It should  be stressed, though,  that their approach
is a \textit{hybrid} of extrinsic and intrinsic
pseudogap theories, where the temperature dependence of the
various gap parameters corresponds to the extrinsic case
(shown in  Fig.~\ref{fig:gaps}b), whereas
the dispersion and superfluid density corresponds to an
intrinsic pseudogap. As shown in this paper, pair-breaking effects on
$T_c$ can be successfully
addressed at a semi-quantitative level both
in intrinsic and extrinsic models\cite{Williams1998}. 
It should be noted that
the rather strikingly different sensitivities of $T^*$ and $T_c$
to impurity concentration
which are found experimentally,
are often taken as an indication that the cuprate pseudogap
cannot be intrinsic, i.e., related 
to the superconductivity, itself.  Indeed similar results are found
in the presence of magnetic field pair-breaking\cite{Kao3} and it should be viewed as
one of the fundamental results of this paper that this inference is incorrect.
\textit{The differences lie in the fact that an 
excitation gap is present when $T_c$
is established, but not at $T^*$, and it is this gap in the density of
states that contributes to  the stronger pair-breaking effects in $T_c$}.  
Indeed, it is precisely this excitation gap which invalidates
the results of conventional AG theory.
%While there is no definitive experiment to distinguish between these
%two schools, we have argued elsewhere\cite{Jelena,Drew} that the
%intrinsic dispersion leads to smaller and more benign modifications of
%BCS theory, which appear to be more
%consistent with current electrodynamic experiments.  
It may be necessary eventually to incorporate an even more
local treatment of pair-breaking than that
discussed here, but \textit{such a Bogoliubov-de Gennes
generalization must include
pseudogap effects}. Indeed, the very basis for a more local
treatment of impurities\cite{Franz1997} is the observed small coherence lengths, which
are at the heart of the present ``intrinsic" pseudogap theories\cite{Kosztin1}.

In summary, in this paper we find
within two diametrically
opposed pseudogap schools, that pseudogap effects  at and 
\textit{below $T_c$} must play an essential role in pair-breaking. 
While there is no definitive experiment to distinguish between
these two schools, we have argued elsewhere\cite{Jelena,Drew}
that the intrinsic dispersion leads to smaller and more benign
modifications of BCS theory.
In both theoretical approaches, the rather robust behavior 
for $T^*$ 
and the associated
excitation gap in the underdoped regime,
found in the presence of impurities may be associated
with the widely observed superconductor-insulator transition\cite{Uchida}.
Superconducting coherence is more readily destroyed than is the
excitation gap (and $T^*$), thereby leading to an insulating
state when $T_c$ is suppressed to zero, in much the same way
as in the presence of applied magnetic fields\cite{Kao3}.
While there are clear differences, seen particularly in
electrodynamical calculations\cite{Jelena,Drew} (as well
as density of states effects) between
the intrinsic and extrinsic pseudogap schools, the impurity
sensitivities of $T_c$ within these two different
approaches are quite similar, and reasonably
consistent with experiment. This similarity derives from
the fact that both mean field theoretic calculations of
$T_c$ have a general BCS-like character, except for the
presence of a (pseudo) gap
at the onset of superconductivity.
For $T^*$ the differences are more apparent in the overdoped
regime and this, in turn, reflects the fact that $T^* \rightarrow 0 $
in one case (extrinsic), whereas $T^* \rightarrow T_c $ in
another (intrinsic).
In this paper we have set the stage for a computation of
transport properties 
which require as an essential input, an understanding of impurity
effects. 
The generalization of AG theory presented here should
help to clarify the important role played by pseudogap effects,
at $T_c$ and their relation to impurity-induced pair-breaking.

We acknowledge very useful conversations with Q.~Chen.
This work was supported by NSF-MRSEC Grant No. DMR-9808595 (YK, JS, AI, KL) and
by NSERC of Canada and Research Corporation (YK).
%\bibliographystyle{apsrev}
%\bibliography{/home/ykao/PAPERS/DIRT/Ref.bib}
%\bibliography{./Ref.bib}

\begin{thebibliography}{38}
\bibitem[{Lif()}]{Lifetime}
\bibinfo{note}{In the intrinsic model, the distinction between the
  superconducting order parameter and the pseudogap below $T_c$ can only be
  observed through the different lifetime of the condensate and the
  finite-momentum excitations. See Ref~\onlinecite{Chen2001}.}
\end{thebibliography}

\end{document}